**Investigating Algorithm Review Boards for Organizational Responsible Artificial Intelligence Governance**

Emily Hadley[1], Alan Blatecky[1], Megan Comfort[1]
[1]RTI International, Durham, NC

**ABSTRACT**

Organizations including companies, nonprofits, governments, and academic institutions are increasingly developing, deploying, and utilizing artificial intelligence (AI) tools. Responsible AI (RAI) governance approaches at organizations have emerged as important mechanisms to address potential AI risks and harms. In this work, we interviewed 17 technical contributors across organization types (Academic, Government, Industry, Nonprofit) and sectors (Finance, Health, Tech, Other) about their experiences with internal RAI governance. Our findings illuminated the variety of organizational definitions of RAI and accompanying internal governance approaches. We summarized the first detailed findings on algorithm review boards (ARBs) and similar review committees in practice, including their membership, scope, and measures of success. We confirmed known robust model governance in finance sectors and revealed extensive algorithm and AI governance with ARB-like review boards in health sectors. Our findings contradict the idea that Institutional Review Boards alone are sufficient for algorithm governance and posit that ARBs are among the more impactful internal RAI governance approaches. Our results suggest that integration with existing internal regulatory approaches and leadership buy-in are among the most important attributes for success and that financial tensions are the greatest challenge to effective organizational RAI. We make a variety of suggestions for how organizational partners can learn from these findings when building their own internal RAI frameworks. We outline future directions for developing and measuring effectiveness of ARBs and other internal RAI governance approaches.

**INTRODUCTION**

Responsible development and deployment of data science and AI approaches requires commitment and action from a variety of relevant parties. Organizations, including but not limited to public and private companies, government agencies, nonprofits, and academic institutions, can play a particularly important role in designing and implementing responsible artificial intelligence (RAI) governance throughout the data science and AI life cycle [19]. Organizational action is acutely important in the United States given the limited landscape of AI law and regulation [23,24,27,40,63,64]. Even without a legal mandate, many organizations have reputational, financial, and mission-aligned motivations for implementing internal governance to support RAI [23,63].

A few organizations have publicly provided details on their internal governance approaches [65–68]. Algorithm review boards (ARBs) or similar committees with oversight of development and deployment of algorithms have been used or proposed as a potential governance approach [5,47,53,63,69]. Other approaches include internal policies and protocols, chief data ethics officer or similar roles, and audits and assessments [19,21,44,46]. Organizations may review what their collaborators and competitors are doing and incorporate them into their own workflows. However, organizations may also not publicize their internal RAI approaches, which makes it hard to understand or appreciate the breadth of approaches and lessons learned.

Additionally, some publicized internal RAI governance efforts are later followed by a quiet or loud dismantling [6,8,25,54]. This has been a particularly noticeable phenomenon among large tech organizations that have announced RAI teams to great fanfare and then later dismantled them [18,49]. This suggests that there are lessons that other companies can learn about what it takes for committee-like internal RAI governance efforts to succeed and what may lead to failure.

In this work, we investigate internal RAI governance practices by interviewing 17 AI practitioners from a variety of organization types and sectors. We seek to answer the question, *"What experiences have technical practitioners involved in the development, use, or regulation of AI had with internal governance approaches that promote the use of RAI?"* We follow this with a more specific question, *"Is an ARB or similar internal committee which reviews and oversees development and deployment of algorithms (including data science and AI tools) effective?"* We also seek to explore the ingredients required for success and the challenges to internal RAI governance approaches.

We find that practitioners widely view RAI as important, although with varied definitions, perspectives, and experiences, even within the same sector. We discover that ARBs and similar boards have been implemented, particularly in finance and health sectors, although these are not standardized or well measured. We provide novel and explicit detail of these ARBs in practice and find that they are among the more utilized and preferred internal governance approaches. We hear that existing Institutional Review Boards (IRBs) are insufficient for organizational management of RAI. We learn that integration with existing approaches and leadership buy-in are among the attributes identified as critical to RAI success and overcoming major hurdles including the often-perceived financial tension of profit with RAI.

We discuss how practitioners, executives, and other organizational actors can use these findings to guide development of their own internal governance strategies including but not limited to ARBs. RAI researchers can review the results to better understand the state of internal governance in practice and regulators can use this work to inform guidelines related to ARBs and other internal governance. Future work may continue to define and standardize the structure, function, and best practices for effective use of ARBs with particular attention to metrics that demonstrate success and effectiveness of ARBs.

**BACKGROUND AND RELATED WORK**

Prior work has established that data science and AI approaches have the potential to provide great benefit, but also the potential to cause substantial harm [31,34,48,50]. Existing governance and regulatory frameworks are generally deemed inadequate or too limited to address these harms [7,11,41,70]. New national and international "soft" and "hard" legal governance frameworks have been proposed and are at varying states of acceptance and implementation [20,56,71,72]. These often center RAI principles and may include consideration of concepts like validity, reliability, safety, security, resiliency, accountability, transparency, explainability, interpretability, privacy, fairness, or mitigation of harmful bias [52]. Yet, organizations, including but not limited to public and private companies, government agencies, nonprofits, and academic institutions, are not waiting for these frameworks to be established and are already developing and deploying AI and data science tools [15,36,38,73]. In this context, organizations may be opening themselves up to substantial reputational and financial risk if they develop or deploy a data science or AI tool that leads to great harm [23,63].

Organizations therefore have a recognized interest in implementing internal governance approaches to manage and mitigate these risks. Existing literature suggests that review boards or committees may be appropriate for internal governance of AI [5,47,53,63,69]. Theoretical proposals have included an AI Research Review Committee, AI Ethics Board, and a Data and AI Ethics Committee [22,47,74]. Some of these proposals touch on topics like proposed membership of a review board, logistics of board meetings, and the scope of board review, although none observe the use of these boards in practice. In this work, we aim to contribute real-world observations to this literature. We use the term "Algorithm Review Board (ARB)" because this type of board or committee is intended to encompass a technical and societal review including but not limited to ethics [12,19,29]. "Algorithm" is also a larger umbrella than AI and recognizes that many organizations may want to use the same review approach for machine learning, data science, and other algorithms, even if not technically considered AI systems.

One notably similar approach is the use of IRBs [1]. These review boards have been proven to better ensure the safety of human subjects in research, and some have proposed IRBs as appropriate for RAI [5]. However, others have noted that the purview of IRBs may lack sufficient scope to adequately assess, document, and mitigate AI risks [22]. In this work, we seek to understand practitioner perspectives on the role, if any, for IRBs in RAI internal governance, and their relationship to ARBs.

Several other promising approaches have been proposed or implemented for organizational RAI governance [19,46]. These may operate in tandem with a review board or committee. One approach is the use of policies, procedures, or protocols to establish a standardized approach to RAI at an organization. Their impact may vary from voluntary suggestions for best practices to exacting definitions and explicit instructions with established consequences for impermissible actions [75–78]. Another approach is granting one individual specific responsibility for reviewing and approving uses of AI, such as a Chief AI Officer or a Chief Data Officer (CDO) [57,58]. A 2022 report found that 56% of more than 850 survey organizations relied exclusively on a CDO or similar role for RAI compliance [79]. A third approach is the anticipation or measurement of potential harms through internal impact assessments or external audits [21,43,51]. These approaches often explore the technical details of a proposed AI solution. Results may be reviewed by an ARB or similar committee or a CDO. A final approach considered in this work is required RAI training for individual contributors [44]. This training may help individuals incorporate RAI best practices directly into their contributions to AI. In this work, we seek to contextualize the use of ARBs in practice with other known approaches.

ARBs and similar RAI approaches may be new for sectors and industries, particularly those that have not historically used enterprise risk management or similar regulatory approaches. The AI change management literature suggests that aligning AI adoption with company values, leadership buy-in, AI education for all staff, and sufficient budget are important for facilitating adoption of AI [16,17,55]. Highly regulated sectors like finance that already use model risk management practices like the "three lines of defense" model may be better positioned to extend existing approaches to AI [51]. In this work, we seek to understand the success and challenges faced by practitioners when implementing RAI in both highly regulated and less regulated sectors.

**METHODS**

To understand the current state of internal governance practices for RAI with a focus on ARBs, we conducted interviews with 17 participants who work with AI or data science across a variety of organization types and sectors.

**Participants**

We utilized a purposive sampling approach with stratified quotas (minimum n = 3) for type of organization (Academic, Government, Industry, Nonprofit) and sector (Finance, Health, Tech, Other) [9]. The objective was to recruit a wide variety of practitioners who have worked with AI and data science in a range of subject areas. We explicitly chose participants who were most likely to actively participate in RAI approaches rather than individuals in leadership or marketing roles who might be familiar with RAI approaches but lack practical experience. Table 1 illustrates the varying organization types and sectors of participants. We recruited our participants through direct contact; attendance at the 2023 Fairness, Accountability, and Transparency Conference and 2023 Joint Statistical Meeting; social media posts (e.g., Slack groups and LinkedIn), and snowball sampling from participants. We contacted 35 individuals and interviewed 17 between July and September 2023.

**Table 1: Participants by Type of Organization and Sector**

|  | Finance | Health | Tech | Other | Total |
|---|---|---|---|---|---|
| **Academic** | 1 | 1 | 1 |  | 3 |
| **Government** |  |  |  | 3 | 3 |
| **Industry** | 2 | 3 | 3 |  | 8 |
| **Nonprofit** |  |  | 1 | 2 | 3 |
| **Total** | 3 | 4 | 5 | 5 | 17 |

All interviewees reported some connection to AI or data science in their work, with AI and data science broadly defined. Seven interviewees described training and building models directly. Four other interviewees discussed using data science approaches that did not necessarily involve training new models. Three interviewees discussed the use of generative AI in their work. Four interviewees work in evaluation or regulatory capacities for data science or AI at their organization or other organizations. Three interviewees are actively involved in research that influences the RAI field broadly.

**Interview Process**

Interviews with participants ranged from 30 minutes to 1 hour and were conducted by video or voice call. Interviews were voluntary, and all results were anonymized for both individual and organizational identifiers. The study and interview protocol were reviewed by the RTI International IRB.

A structured interview instrument was developed using the Interview Protocol Refinement (IPR) framework [10]. Questions were developed and revised iteratively as study interviews progressed. The interview protocol was piloted with two RTI International data scientists and further refined based on feedback (Appendix A).

**Qualitative Analysis**

The Iterative Categorization technique was utilized for analyzing the qualitative data [35]. Deductive codes were derived from the structured interview instrument. Inductive codes, often subcodes to deductive categories, were added during analysis (Appendix B). In the first phase of analysis, coding extracts were summarized, reviewed, and grouped along with qualifiers and identifiers that linked to the source material. In the second phase of analysis, each analysis file was read and re-read with attention to patterns and contradictions. These were assessed considering participant attributes, namely organization type and sector. Summary documents were linked together to form the basis of the findings.

**FINDINGS**

Findings were grouped into three main areas: *Individual Context and Perspectives on RAI* with descriptions of broad participant views on RAI; *ARBs and Other Internal Governance Practices* with an overview of participant experiences with RAI governance; and *Ingredients for and Challenges to Success of Internal RAI Governance* with summaries of the attributes that participants described as important for or challenges to the success of RAI approaches at organizations.

**Individual Context and Perspectives on RAI**

All interview participants were familiar with and able to define RAI. Many participants (n=12) stated a definition that described preventing harm, discrimination, or bias in the practice of developing and deploying AI tools. Around a third (n=6) of participants used language describing RAI in terms of providing a meaningful benefit to customers or society. Notably, a small number of participants (n=3) considered that RAI may be an ambiguous concept without a clear definition and instead is a term that serves as a buzzword or corporate marketing ploy.

A subset of participants (n=3) extended the discussion of the definition of RAI to sociopolitical implications of AI development. One participant considered how AI deployment could perpetuate historical power imbalances reminiscent of colonialism and imperialism, especially when well-resourced nations implement AI technologies in the Global South. Others (n=2) critiqued the capitalist drive underlying AI development, suggesting it could misshape AI uses by favoring profit over societal good and overlooking non-lucrative yet socially vital AI applications.

The importance of RAI was not asked directly, yet nearly every participant (n=16) provided a perspective on the rationale for why RAI was important to their work. Seven individuals said that RAI was crucial for developing tools that were trustworthy products, including all individuals working in the health care sector. A minority of participants (n=3) cited an ethical imperative to using RAI practices that supported a moral work culture which allowed them to sleep at night, have peace of mind, or anticipate a future where they will not regret their work on AI. Six individuals described how RAI is important for fostering conversations on the limitations of AI development and use broadly. This included amplifying the limitations of a "move fast and break things" tech culture, such as the lack of risk management approaches, misalignment with existing regulations and laws (e.g., Americans with Disabilities Act), and broader concerns of the experiential and qualification deficiencies of individuals and companies receiving funding to develop AI tools.

Some participants contextualized the need for RAI among larger external efforts. Approximately a third of interviewees (n=6) noted that government regulations may impact enterprise RAI, but that there is not yet much guidance for organizations in the

United States. Five individuals called for the federal government to take more action, though two participants noted that even if regulation was implemented, it may not be sufficient or robust enough to ensure RAI in practice. Individuals cited other sectors like cybersecurity (n=3) and consumer finance (n=3) as influences for organizational and government RAI guidance.

**ARBs and Other Internal Governance Practices**

Eight interviewees reported use of a review board for organizational algorithm, data science, or AI governance, including all individuals working in the health sector and finance industry. These boards typically had custom names that would violate privacy protocols if published. Though none were officially called an ARB, they met the definition of an ARB and will be referred to as ARBs in this manuscript. Table 2 provides an overview of the details from participants for the organizations that reported using an ARB as part of an internal RAI governance framework. Seven interviewees who did not report current use of an ARB said that an ARB would be a promising approach in their organizations, including three interviewees working in the tech sector. Only one interviewee said that an ARB was not a promising approach and cited the small size of their organization.

**Table 2: Details of Algorithm Review Boards in Practice**

| | 1 | 2 | 3 | 4 | 5 | 6 | 7 | 8 |
|---|---|---|---|---|---|---|---|---|
| Sector | Government | Tech | Health | Health | Health | Health | Finance | Finance |
| Persons Involved in Board | Privacy officers, researchers, industry, and nonprofit representatives | At minimum one person from a sociotechnical background and one person from a technical background | One-third of board from leadership, such as lead data scientist, manager, or director | Individuals with expertise in ethics, privacy, security, legal, operational, and clinical areas | At minimum, a regulatory expert, a bio-statistician, and operational leader; others as needed, such as a health equity expert or the IRB chair | Clinicians | Model risk officers, with insight from data scientists as needed | Model risk officers, with insight from data scientists as needed |
| Selection Process | Voluntary | Voluntary | Voluntary | Voluntary | Job Requirement | Voluntary | Job Requirement | Job Requirement |
| Scope of Review | Risk-based Review | Review of Documentation | Review Everything | Review of Documentation | Review Everything | Review as Requested in Development | Risk-Based Review of Everything | Risk-Based Review of Everything |
| Use of IRB | Not Applicable | Separately Used | Not Specified | Separately Used | Required by Board | Not Applicable | Not Applicable | Not Applicable |
| Meeting Cadence | Quarterly | Ad Hoc in Release Process | Ad Hoc in Release Process | Monthly | Monthly; Async & Ad Hoc In Release Process | Async | Async | Async |
| Decision Power | Board has no decision power | Board has no decision power | Majority vote | Unanimous vote | Co-Chairs | Board has no decision power | Consensus Approval | Consensus Approval |
| Measure(s) of Success | Not specified | To be determined | N approvals; N issues caught before deployment | To be determined | N registered tools; N reviews; outcomes of reviews | Not specified | N past due findings needing resolution | Lack of public failures as measure of success |

Table 2 highlights the wide variety of expertise and contributory roles required for each ARB. Membership combinations included privacy officers, regulatory experts, biostatisticians, and persons with sociotechnical and technical expertise. Some participants (n=2) at companies with larger ARBs of 12 or more people described the use of subcommittees with three or four individuals to delegate reviews. Slightly more (n=5) boards solicited volunteers for participation on ARBs while the remaining (n=3) included board participation as a job role or requirement for specific positions. Among those with voluntary ARBs, experiences varied with two individuals reporting that more individuals volunteered than were needed while another individual said that it was challenging to ensure timely contributions from volunteers since the review board was secondary to their main job priorities.

The extent and rigor of algorithm review processes vary considerably among ARBs in Table 2. In the finance sector and some of the health sector, every algorithm and model undergo scrutiny by the ARB. ARBs in other sectors report only reviewing documentation or reviewing by request of developers. Three individuals reported that the intensity of these reviews is calibrated to the potential risks involved with higher risks necessitating more stringent reviews. ARB review procedures may involve a scoring system or threshold requirements to guide go/no-go decisions. For some ARBs, this process is iterative with reviews at multiple development checkpoints which allows for committee feedback to developers for model improvement or conditional implementation. Furthermore, several individuals emphasized the importance of ARB involvement from the planning stage, highlighting a proactive approach to governance that aligns with the lifecycle of technology development.

As noted in Table 2, three individuals reported use of an IRB and one individual in the health sector described it as a separate requirement in the RAI process. No individual reported IRBs as sufficient or appropriate for ensuring responsible use of AI. Six individuals reported at least one limitation to using IRBs as part of an AI internal governance process. These limitations included:

- The scope of an IRB is limited to research and generally not required or well known in industry settings.
- Data projects are often exempt from IRB review, impacting their usefulness in an internal AI governance process.
- Unpopularity and cultural backlash can result from the perception of IRBs as barriers to expediency and productivity.
- The conception and principles of an IRB are North American–centric and not globally known, accepted, or implemented.
- IRBs are not designed for long-term monitoring required for deployment of AI.

ARB meeting logistics also varied. Four individuals, all from the finance and health industries, described a mostly asynchronous review board process involving review of documentation and technical reports with meetings among a smaller operational team as needed. Some individuals (n=3) described having review board meetings scheduled to align with the release process. One participant described requiring all algorithms to be presented in a setting that was also open to the broader data science community within their organization. A second participant described previously requiring all developers to present their RAI reports to the ARB, but then moving to an asynchronous process when this became too burdensome and challenging to coordinate with calendars. Time for approval varied. One finance interviewee said low-risk approvals may be nearly instant while more complex approvals may take 2 weeks to 1 month. One health interviewee said the full approval process may take over a year with multiple 3- to 7-week checkpoints if deployed in a setting with electronic health records. Five individuals reported that their internal governance committees were empowered to make go/no-go decisions about the continued development or deployment of an algorithm. Two of these individuals reported that the decision was made through voting.

Defining success of an internal governance board was a challenge for nearly all interviewees. One interviewee explicitly said that this was challenging because there was no clear industry understanding of success. Some interviewees (n=4) from health and finance sectors reported tracking logistical metrics like the number of reviews, approvals, tools registered, and the percentage of deployed tools that had completed quarterly or annual monitoring reviews, but no participant provided a clear threshold for what was considered successful. Two individuals from health and finance sectors reported a desire to attempt to track the number of RAI-related bugs that were identified or mistakes that were caught in the review process and better quantify anecdotal reports of these items. Two individuals from health industries mentioned gathering feedback from developers and reviewers about the internal review board process, and one mentioned wanting to design a survey to better collect this information. One individual considered strength of brand reputation and lack of being sued as potential metrics of success.

Table 3 details other RAI approaches used at participant organizations. Written policies, procedures, and protocols were the most common with every organization reporting an ARB also having a RAI policy, procedure, or protocol. These were reported as foundational to company culture and an important precursor to an ARB. ARBs often relied on procedures and protocols in defining the review process (n=4). These policies were sometimes wrapped into existing data governance policies (n=2). Some individuals (n=2) reported policies as informal and not binding. Policies, procedures, and protocols on AI governance, if any, were generally only available internally to company staff. The exception was government interviewees who all described efforts to proactively publish public protocols which, while voluntary, were considered best practice. Internally, interviewees reported using tools such as Slack channels, wikis, GitHub repositories, and risk management databases to disseminate information. Yet interviewees still expressed concern about challenges for communicating policies and procedures (n=7) and potential limits of their impact without accompanying governance mechanisms (n=4).

**Table 3: Other Internal Governance Approaches**

|    | Sector     | Algorithm Review Board or Similar | Policy, Procedure, or Protocol | Individual with Responsibility | Impact Assessment or External Audit | Training or Certification for Developers |
|----|------------|-----------------------------------|--------------------------------|--------------------------------|-------------------------------------|------------------------------------------|
| 1  | Government | X                                 | X                              | X                              |                                     |                                          |
| 2  | Nonprofit  |                                   | X                              |                                |                                     |                                          |
| 3  | Nonprofit  |                                   | X                              |                                |                                     |                                          |
| 4  | Tech       | X                                 | X                              |                                |                                     |                                          |
| 5  | Tech       |                                   |                                |                                |                                     |                                          |
| 6  | Health     | X                                 | X                              |                                |                                     |                                          |
| 7  | Nonprofit  |                                   |                                |                                |                                     |                                          |
| 8  | Academic   |                                   |                                |                                |                                     |                                          |
| 9  | Health     | X                                 | X                              |                                |                                     | In Development                           |
| 10 | Government |                                   |                                |                                |                                     |                                          |
| 11 | Health     | X                                 | X                              |                                |                                     |                                          |
| 12 | Government |                                   | X                              |                                |                                     |                                          |
| 13 | Health     | X                                 | X                              |                                |                                     |                                          |
| 14 | Academic   |                                   | X                              |                                |                                     |                                          |
| 15 | Tech       |                                   | X                              | X                              |                                     |                                          |
| 16 | Finance    | X                                 | X                              | X                              | X                                   |                                          |
| 17 | Finance    | X                                 | X                              | X                              | X                                   |                                          |
|    | Total      | 8                                 | 13                             | 4                              | 2                                   | 1                                        |

Four individuals, including three at organizations with ARBs, reported having an individual with responsibility and decision-making power for approval of algorithms. Titles were often similar to Chief AI Officer or CDO. The main benefit reported for this RAI approach was the speed with which decisions could be made and clear responsibility in the chain of command. However, several limitations were noted of having a singular individual with decision power including lack of broad knowledge, too much work, potential conflict of interest, influence of personality, lack of diverse opinions, and disruption following employee turnover. An ARB or committee structure was described as a preferable alternative as the use of a diverse group of experts for review addresses many of these concerns.

External audits, another promising RAI approach, were only reported in active use by individuals in the financial sector. While a few other interviewees (n=3) said that these may be a useful approach, they were joined by seven other individuals in identifying reasons that audits or assessments were hard to implement:

- Undefined criteria for what constitute an effective audit (n = 4).
- Company belief that audits are not needed (n = 3).
- Expensive to perform an audit (n = 2).
- Logistically not ready for implementing audits (n = 2).
- Pushback against private sector audits (n = 2).
- Audit paperwork is not useful when buried through legal privilege (n = 1).

Only one interviewee reported that their organization was actively developing employee training related to RAI. Ten individuals said that RAI training would be useful, but seven also noted substantial limitations. These included concerns that training is often superficial (n=3), RAI is still a developing space and there is not agreement on what a training would look like (n=2), and pushback against more employee trainings unless directly relevant to an employee's work (n=4).

*Ingredients for and Challenges to Success of Internal RAI Governance*

When asked about what was needed for successful use of RAI internal governance, interviewees cited several factors shown in Table 4. The most frequently mentioned item was integration with existing institutional bodies or methods. Interviewees from the finance sector described how algorithm review and an ARB-like committee fit well with existing governance mechanisms in a risk management framework. One interviewee emphasized that large companies already have functions like legal audit, enterprise risk management, and privacy and security practices, and that internal regulation of AI and data science should use the same governance approaches. A second important factor was leadership buy-in. A few (n=3) participants considered this the most important factor for the success of RAI internal governance. They described how authentic commitment of executive leaders within an organization to RAI was often accompanied by staff, funding, and attention. One interviewee described it as, paraphrased, "If your boss cares, you care." Another noted that a single data scientist was not going to be successful implementing a scalable RAI framework. Other common important factors like high contributor and developer receptivity to RAI and existing regulatory culture were noted as intertwined with leadership and organizational commitment to governance and regulations broadly. Human elements including transparent use of AI systems, education of staff, and required human review for high-risk decisions were also noted as important to the success of RAI governance.

**Table 4: Attributes for RAI Success**

| Attribute | N Reporting |
|---|---|
| Integration with existing institutional governance bodies or methods | 10 |
| Leadership buy-in | 6 |
| High contributor and developer receptivity | 6 |
| Existing regulatory culture | 6 |
| Transparent use and disclosure of AI systems | 4 |
| Education of staff | 4 |
| Required human-in-the-loop review for high-risk decisions | 2 |

Interviewees also raised several potential challenges (Table 5) to the success of internal RAI governance. The most common issue (n=13) was financial tensions between RAI and profitability. Over a third (n=7) of individuals described how RAI governance was seen as losing money for an organization because it added friction to or slowed down development. Some participants (n=3) described rapid deployment as a higher organizational priority than ethics. A few participants (n=3) expressed concerns that in economic downturns, internal governance structures and roles appear vulnerable, and that internal RAI governance may not be financially viable for small to medium sized organizations that are unlikely to see substantial profits from AI. Contrasting this perspective, a few individuals (n=2) from finance and healthcare organizations stressed that patient and customer well-being is paramount, especially for maintaining an organization's reputation, and expressed that their organization would not compromise on RAI governance even when it was expensive.

Participants also highlighted several other challenges (Table 5). General challenges were educating staff on existing and new internal RAI efforts, particularly when announcements through email or messaging apps are ignored, and addressing the perception that governance is unnecessary bureaucracy. For larger organizations with existing AI needs, challenges included ongoing monitoring of deployed AI and integrating RAI into vendor and procurement processes. Vendor issues included a lack of standardized RAI reporting as well as unanswered questions about legal liability in the case of AI harm. Enforcement of responsibility was seen as a major challenge, especially at organizations where the ARB lacked go/no go decision power for use of algorithms. Large scale projects and the need to work fast were seen as potential issues for RAI approaches, including an ARB, which often work more slowly. A few individuals noted that it is relatively easy for RAI governance to be misused as a marketing tool or performative action without substance and that it is hard to futureproof RAI governance to be adaptable for unknown technologies and uses. One interviewee noted that even when a company hires RAI staff with a grand vision, they must have a budget, real authority, and realistic plans, or else they will fail to have an impact. Another challenge was setting clear expectations for what successful RAI governance looks like due to the invisibility of preventative success. One individual described how new employee performance metrics were needed because RAI contributors and teams were often among the first to be fired as they did not appear to perform well on existing code or efficiency-based performance metrics. Finally, company size affects RAI governance, with smaller entities struggling with resources and larger ones battling bureaucracy and siloed departments.

**Table 5: Challenges to RAI Success**

| Challenge | N Reporting |
|---|---|
| Financial Tensions | 13 |
| Educating Staff on Policies | 7 |
| Introducing More Bureaucracy | 7 |
| Ongoing Monitoring | 7 |
| Vendor and Procurement Challenges | 7 |
| Enforcement of Responsibility | 6 |
| Scale or Speed of Work | 5 |
| RAI as Marketing or Performative | 4 |
| Futureproofing | 3 |
| Expectation Setting | 3 |
| Size of Company | 3 |

**DISCUSSION**

The findings detail the wide range of experiences that participants have had with RAI governance across organization types and sectors. In this section, we discuss these findings and their implications, particularly for relevant parties seeking to implement an ARB or other internal RAI approach.

**AI practitioners view RAI as important, though with varied definitions and perspectives.**

Nearly every participant considered RAI important to their work, regardless of organization type and sector. Over a third of participants specifically said that RAI was helpful for changing the tone of the conversation around AI away from the "move fast and break things" culture in which modern internet technology was cultivated and toward a more intentional, inclusive, risk-averse ethos. This suggests that participant experiences in developing and using AI and data science tools align with the broader literature that also recognizes RAI as important and necessary for a break from historical AI development norms that have permitted the creation and deployment of harmful AI tools [26,32]. A few participants further situated their RAI perspective within the context of colonialism, imperialism, and capitalism, topics that have also been considered specifically within RAI literature [2,33,37]. Practitioners with these perspectives may be important contributors to internal RAI governance approaches because they may better recognize and address colonial and imperialistic undertones in AI applications.

When asked to define RAI, interviewees shared some overlapping concepts focused on preventing harm and providing benefit with AI and data science, but there was no singular, standardized definition. This also aligns with literature that has suggested a variety of ways to define RAI and attempts to distinguish RAI from ethical or trustworthy AI [13,26,45]. Participants also shared that government regulation was needed but organizations can and do act on RAI sooner than laws, rules, or other guidance. Practitioners working on RAI governance for organizations can take away that organizations are already implementing internal RAI governance approaches without waiting for a standardized definition of RAI or government regulation.

**ARB-like boards have been implemented, particularly in finance and health sectors, though not standardized or well-measured.**

Nearly half of organizations (n = 8) reported having an ARB-like board or committee tasked with reviewing AI or algorithms, although most were custom-named, and none were called an ARB. This reflects a shared perception of the importance of a review board or committee but also the lack of standardization. Health and finance sectors seemed to have more robust review boards that were also more likely to require review of all models and algorithms. Although both sectors are known for strong quality management approaches and rigorous risk management frameworks, the robust nature of algorithm review with boards and committees in health care specifically has not been previously reported in literature. Organizations in health and finance sectors that use AI without RAI governance should be aware of and likely seek to emulate the peers described in this work.

Most ARBs require a mix of technical and nontechnical experts, but requirements varied from privacy to ethics to IRB to sociotechnical expertise. This aligns with literature that suggests that review boards should likely be tailored to specific sector and organization needs [22,74]. Slightly more ARBs had voluntary participation, but those with participation as part of the job requirement described this generally as positive. Organizations considering development of an ARB should be intentional in deciding what skillsets are needed and how participation is determined.

ARBs also varied in degree of algorithm or AI review. Some ARBs, especially in health care and finance, reviewed all models and algorithms, requiring calculation of various measurements, detailed documentation, and other organization-specific items. Other organizations only required review of the high-risk deployments while still others reviewed only documentation or only did review as requested by developers. These findings suggest that organizations are still figuring out what works best, and that what works well for one organization may not be appropriate for another. Most ARBs either met asynchronously or ad hoc with a smaller subset of the team, suggesting that the routine logistics did not require frequent meetings. Even ARBs that did meet regularly met monthly or quarterly rather than weekly or even more often. This suggests that frequent meetings are not a hallmark of success for ARBs, and strong logistics that support asynchronous work with smaller and less frequent meetings may be preferable.

The strategy for using ARBs to make decisions also varied. Health and finance generally granted boards go/no-go decision power while the sole government and tech examples did not grant decision-making power to the board. This may reflect differences in corporate culture and defining a committee as advisory versus regulatory. The decision process varied among ARBs with decision-making power, with the spectrum ranging from consensus and unanimous decision-making to majority votes to decisions made by board co-chairs. This again reflects the lack of standardization in ARBs with choices that likely align with existing culture and processes. Future work may explore which specific attributes are most effective in particular scenarios. Organizations seeking to develop their own RAI internal governance approaches can assess the variety of decisions made in this work and select attributes that may be the most successful in their own working environments.

Nearly every participant at a company with an ARB struggled to define a measure of success and often focused on the quantitative measures that were easiest to capture like number of registered tools, number of issues caught, and number of past due findings. Lack of ability to measure success may translate to a lack of ability to prove success, linked with other participant concerns that RAI individuals and teams are often the first to be fired because of an inability to convey impact. In addition, the success of a RAI individual or team is shown in lack of major failure or effort, which is harder to measure and communicate than obvious and often harmful AI incidents. Several participants expressed the need for better, clearer, and more significant metrics for ARBs. Organizations seeking to implement RAI approaches should be aware of this challenge and address it up front when hiring RAI staff and developing ARBs or similar. Researchers in the RAI space can seek to meaningfully contribute by suggesting alternative measures for individual and team contributors that are useful in a job evaluation environment.

**IRBs are not sufficient for organizational management of RAI.**

Limited previous literature has theorized that IRBs may be good governance tools for RAI; however, this work strongly suggests that IRBs are not appropriate alone for ensuring RAI. More than half of participants discussed the limitations of IRBs, including their limited scope, unpopularity, North American–centricity, and lack of use for long-term monitoring beyond human subject research. One participant in a health care setting described how an IRB review was required, but this was entirely separate from the review by the ARB-like board. Individuals designing an internal AI governance program at an organization with an IRB should confer with the IRB to discuss RAI resources needed for research with human subjects within IRB purview but should acknowledge that use of an IRB is not sufficient for RAI and other mechanisms such as an ARB are considerably more comprehensive and robust.

**ARBs are a strong contender among varying RAI approaches.**

Although ARBs were the focus of this investigation, we recognized that they were a novel concept and wanted to also hear about other RAI approaches organizations had attempted or were using. RAI policies and protocols had higher adoption than boards, but participants also expressed substantial concern about meaningful impact, especially when they were voluntary or unaccompanied by other RAI efforts. This aligns with existing literature that suggests that companies may use policies and protocols for "ethics washing" and to give the appearance of caring about RAI, even if it is not a real effect [4,28,42]. ARBs may be more impactful than policies or procedures alone because they may better enforce accountability. Four individuals reported having a single decisionmaker with some degree of RAI decision-making responsibility, but more individuals described limitations of this approach, including not being future proof, lack of power or responsibility, lack of diverse opinions or broad knowledge, personality, conflict of interest, and too much work. Enterprise model risk management has historically suggested that a single decisionmaker with real responsibility can be a critical part of a risk management framework [51], but participant concerns suggest that this approach may not be appropriate for RAI.

Audits were of substantial interest to participants, although only interviewees from the financial sector reported real-world experience with these. Participants also had numerous questions on RAI audit implementation like what criteria would constitute an effective audit, who pays for it, review logistics, pushback against private sector audits, and concern that the audit materials would not lead to anything if buried under legal privilege. A variety of audits, assessments, and associated certifications are in development [21,43,59,61,80], and it would be worthwhile for individuals contributing to the development of audits to consider the concerns raised by practitioners in this work. Similarly, although only one organization in this work is actively exploring RAI training for employees, many individuals were interested and said that if well-designed and relevant, they could become an important part of the employee workflow. Overall, ARBs appeared to be the most actively used and impactful RAI governance approach. Future work with audits and employee training may be useful to organizational partners.

**Integration with existing approaches and leadership buy-in are among the attributes critical to RAI success.**

The most cited attribute for successful implementation of RAI internal governance was integration with existing organizational regulatory approaches. This was particularly important in sectors like health and finance that already have a strong regulatory culture, which may be lacking in other sectors like tech. Organizations considering implementation of RAI should consider first inventorying existing data governance approaches at their organization such as responsible use policies, data use agreements, or data governance committees, and integrating novel RAI approaches with these existing ones. This reduces the labor and infrastructure burden for introducing RAI approaches as compared to starting from scratch and may also increase buy-in from relevant partners.

The second most cited attribute, and for some individuals the most important, was leadership buy-in. This similarly aligns with change management literature that suggests that leadership buy-in is crucial to driving large-scale change at a company [16,17]. Leadership buy-in comes with attention and often additional resources like staff or funding. However, leadership buy-in must be authentic. Executives and other contributors in leadership roles should seek to educate themselves on RAI and to meaningfully leverage their power to fund RAI work and staff RAI positions. Although the RAI space is not yet entirely defined, executives can lead the way with these early investments and set the tone for a cultural change.

Other factors that individuals interested in implementing internal RAI approaches should pay attention to include transparency, disclosure, and a requirement for human-in-the-loop review. These are seen as important for developing and maintaining trust, especially in high-risk or sensitive sectors like criminal justice. Indeed, recent guidelines like the landmark US Executive Order on Safe, Secure, and Trustworthy Development and Use of Artificial Intelligence and the provisional European Union Artificial Intelligence Act have called for attention to these very principles [20,56]. Even organizations not subject to these guidelines should consider taking steps to align with these human-centered attributes given their importance for RAI success.

**Financial tensions and other concerns present major challenges to RAI implementation.**

Many participants expressed concern that profit was at odds with safe, scalable, and reliable tools, and that this financial tension was a major challenge for the success of RAI. This concern echoes concerns among financial tensions related to RAI as seen in literature and the news media where companies have reported that ignoring RAI could result in substantial losses, but that the upfront RAI costs may be difficult to meet [3,39,60,62]. Two participants from the health and finance sectors were notable exceptions in saying that they did not observe financial tension with RAI because of the focus on the patient or customer and the importance of reputation. Organizations seeking to implement internal RAI approaches should consider financial tensions up front and seek alignment with organizational mission, vision, reputation, or values. Very real labor, time, and resource costs exist for implementing RAI, and this burden may be particularly large at small organizations. Organizational leaders should consider proactive methods to address these costs, such as obtaining explicit funding for RAI approaches, justifying costs to clients, and seeking effective implementation methods that will minimize costs where possible.

Another major challenge was educating staff on RAI policies and protocols. This is a recognized issue in the change management space where simply creating a policy is not enough; it must be communicated and enforced to have impact [16]. Organizational leaders seeking to implement RAI governance should work closely with internal communications teams to explore approaches to best communicate and enforce RAI approaches. Suggestions may include additions to new hire onboarding, annual trainings, and presentations for internal contributors. Working groups, blogs and newsletters, and internal communication platforms may also play an important role. Education is also important for framing the context in which RAI is implemented. Interested organizations should be aware that individual contributors may see RAI approaches as simply another layer of bureaucracy and can use these same educational approaches to help individual contributors appreciate the larger ethical and regulatory motivations.

Organizational leadership should also consider a handful of specific challenges. One is ongoing RAI monitoring of deployed tools. Organizations should consider developing a maintenance plan and obtain funding prior to deployment. Ideally this framework is flexible to novel future use cases that are not yet known. Another consideration is how organizations that work with vendors may integrate RAI with existing procurement protocols. The legal risks of AI are not yet well defined, particularly when it comes to responsibilities that emerge from organizational relationships. Organizations should consider holding vendors to the same RAI standards as internal tools. Vendors and procurement are open areas of development, and more standardized documentation and resources may become available in the coming years [14,30,81,82]. A final consideration is organizational cultural changes such as developing a culture of responsibility to meaningfully hold individuals accountable and combat a culture where RAI is performative or marketing. It is easy for organizations to commit to RAI in name only.

**Limitations**

A few limitations exist for these findings. First, the use of a purposive sampling approach means that these results are from an intentionally narrow sample and are not intended to be broadly generalizable. Second, participants were deliberately selected to be more technical or task-oriented contributors and therefore may not have a broad or landscape view of the full RAI governance landscape at an organization. This is a particular issue at large organizations where individuals may not have awareness of RAI approaches upstream or downstream of their contributions. Finally, interviewees were conducted in summer 2023, which preceded the US AI Executive Order in October 2023 and provisional EU AI Act in December 2023. These regulations may influence the RAI approaches described in this text, particularly those shared by individuals affiliated with government institutions or that operate in the EU.

**Implications and Future Work**

Organizations considering implementing internal RAI approaches can use the findings from this work to inform their own decisions. Specifically, organizations considering an ARB can use these results to inform development of their own ARB. Individual contributors and organizational leaders can also learn from that successes and challenges that participants shared. Government regulators can use these findings to inform rules and regulations related to internal RAI approaches, particularly ARBs or similar committees for review of algorithms.

One major area of future work is defining metrics of success for ARBs. No participant felt entirely confident in their own metrics, and multiple participants mentioned concerns that insufficient metrics may lead to RAI individual contributors or teams continuing to be fired because they do not typically perform as well on existing employment metrics. Other future work may more directly measure which RAI approaches are more successful than others with explicit consideration of effectiveness by company size. This is particularly important for small organizations that lack sufficient resources for robust RAI implementation.

**CONCLUSION**

Organizations are not waiting for AI legislation and are already implementing internal RAI governance approaches, reflecting the importance of RAI governance for addressing the risks of AI. ARB-like boards are already in use, particularly in health and finance sectors. Numerous differences exist in review board structure, scope, logistics, and decision power, suggesting ad hoc or customized development. Other RAI approaches including policies and procedures, CDOs, and audits and assessments are used in tandem with ARBs, although ARBs appear to be more popular and potentially impactful approaches. IRBs alone were considered insufficient for ensuring RAI development and deployment. Integration with existing processes, particularly in sectors like health and finance with strong regulatory cultures, and leadership buy-in were considered critical to the success of internal RAI governance. Financial tensions between profit and RAI cost were a major concern for many participants. Future work should explore measurements of success for ARBs and similar review boards and consider quantitatively comparing ARBs with other RAI internal governance approaches.

**ACKNOWLEDGMENTS**
The authors would like to thank the anonymous participants who generously contributed their time and perspectives. The authors would also like to thank Stephanie Hawkins and Jamia Bachrach for their insights.

**Ethical Considerations Statement**
This research is considered research with human subjects and was reviewed by the RTI International IRB. Participants were provided background information on the study and informed at the start of each interview that they were free to decline to answer any questions, take a break, or stop the interview at any time. Participants also were asked to provide consent for recording of interviews. Stringent confidentiality measures were used to ensure that participants felt comfortable responding freely. Names were only used in correspondence, scheduling, and contact information files. Interview recordings and notes used a pseudonym. The name of the organization that each participant was affiliated with was not collected. Individuals were told they would not be directly quoted, and all quotes would be paraphrased in the final research product. Interview recordings were stored in a secure RTI International storage system accessible only to the research team and destroyed following the completion of the project.

**Positionality Statement**
All three co-authors are staff at a nonprofit research institute. Our experiences as technical contributors and knowledge of the impact a developer can have on a final product may have biased us towards a focus on interviewing individuals with technical roles rather than individuals in leadership or compliance. We also have undergraduate and graduate education in statistics, data science, ethics, and sociology that influenced the direction of this research and provided us with relevant connections for identifying participants.

**Adverse Impacts Statement**
The findings of this work may be used to develop internal RAI governance practices at organizations. Implementing one or more of the suggested approaches may require substantial financial investment, time, or personnel, which may be overwhelming for under-resourced organizations. The suggestions in this work are not guaranteed to align with regulations like the EU AI Act or US AI Executive Order, and organizations may be adversely impacted if they implement suggestions in this work with an aim of compliance with national and international regulations.

**APPENDIX A**
**External Interview Instrument for "Internal Governance Approaches for Promoting Responsible AI"**
**Introduction:** *Thanks for taking the time to speak with me (us) about responsible AI and internal governance today. My name is XXXXXXX and I am a co-investigator for this research. During the interview, we will ask you to share your perspectives on internal governance approaches used by organizations to promote responsible AI, including a few hypothetical scenarios. You are free to decline to answer any questions, take a break, or stop the interview at any time. While we may be discussing experiences related to your current or past work, please do not share any confidential or proprietary information.*

*To minimize the risks to confidentiality, your name will only be used in our correspondence, scheduling, and our contact information file. Interview recordings and notes will use a pseudonym. We will not collect the name of your organization, but we will ask you to select a descriptor for the type of organization (Industry, Academic, Nonprofit, Government, Other) and the subject area you most closely work on (Health, Finance, Technology, Other). We may share aggregated findings related to these areas. Your contact information and interview recording will be stored in a secure storage system accessible only to our research team and destroyed following the completion of our project.*

**Do you have any questions before we begin?**
**Are you okay with us recording this interview?**

Part 1: *To start with, we'd like to ask a few background questions.*
- **What is the best descriptor for the type of organization you work for: industry, nonprofit, academic, government, or other?**
- **What is the best descriptor of the subject area you work in: health, finance, tech, or other?**
- **How would you briefly describe the ways in which you work with AI or data science?**
- **In a few sentences, what does responsible AI mean to you? Feel free to think it through**
    - (*If candidate has not heard of the term, provide a definition*) **Definition**: *Responsible AI generally means developing and using AI in line with considerations of fairness, accountability, privacy, explainability, and societal impact.*

Part 2: *Next, we'd like to ask a few questions about your experiences with internal AI governance.*
- **What internal governance approaches, if any, has your organization tried or actively used to support responsible AI?**
    - *If prompted for examples:* **algorithm review board or data ethics committee, data ethics officer, algorithm audits, AI certifications)**
- I*f interviewee provides examples, continue conversation for each example:*
    - **Can you tell me more about how this approach works in practice?**
        - If in current use:
            - **What has been the impact of this approach?**
            - **What would you say has worked well?**
            - **What would you say hasn't worked well?**
            - **What would you recommend to another organization considering this type of approach?**
        - If not in current use:
            - **Why do you think this approach is no longer used?**
            - **What could have made this approach successful?**

Part 3: *Next, we'd like to ask you a few questions about hypothetical AI governance approaches:*
- *If "algorithm review board" was not mentioned in Part 2:* **Are you familiar with the concept of an Algorithm Review Board?**
    - (*If candidate has not heard of the term, provide a definition*) **Definition**: *A diverse team of experts with technical and ethical expertise that can review and approve AI projects in line with principles such as respect for persons and benefit to society.*
    - **How promising do you think this approach is for your work environment? Feel free to think out loud**
- **How do the following internal AI governance options compare with an algorithm review board?**
    - **Individual with oversight**
    - **Individual ethics training or certification of developers**
    - **External audits or internal assessments**
- **What do you think is needed to support a culture of internal governance for responsible AI?**

o   *If prompted, follow up:* **Alternatively, what do you think is preventing the success of responsible AI internal governance?**
o   I*f financial considerations are mentioned:* **You mentioned financial considerations – can you comment more on how companies do or don't profit on responsible AI?**

*Part 4: Thanks for these responses.*

- **Is there anything related to internal governance and responsible AI that we haven't discussed that you would like to comment on?**

*Thanks for taking the time to participate in this interview. We plan to review the results over the coming months and plan to summarize our findings and submit a publication. Feel free to reach out with any questions or concerns that may come up. Have a good day.*

**Appendix B Qualitative Codes**
**00 BROAD USE OF AI**
**01 RESPONSIBLE AI PERSPECTIVES**
      **01.1 DEFINING REPONSIBLE AI**
      **01.2 PERSPECTIVE ON IMPORTANCE OF RESPONSIBLE AI**
      **01.3 COLONIALISM, IMPERIALISM, AND CAPITALISM**
      **01.4 OTHER SECTORS AS A MODEL**
      **01.5 GOVERNMENT REGULATIONS**
**02 IN-USE PRACTICES**
      **02.0 TIMELINE OF IMPLEMENTATION**
      **02.1 ARB DESCRIPTION**
            **02.1.1 MEETING LOGISTICS**
            **02.1.2 DECISION POWER**
            **02.1.3 MAKEUP OF BOARD**
            **02.1.4 SCOPE OF BOARD**
            **02.1.5 RELATION OF BOARD TO IRB**
            **02.1.6 MEASURES OF SUCCESS**
      **02.2 POLICIES, PROCEDURES, AND PROTOCOLS**
      **02.3 INDIVIDUAL WITH RESPONSIBILITY**
      **02.4 IMPACT ASSESSMENTS AND EXTERNAL AUDITS**
      **02.5 TRAINING AND CERTIFICATION FOR DEVELOPERS**
**03 INGREDIANTS FOR SUCCESS**
      **03.1 INCORPORATION INTO EXISTING INTERNAL BODIES**
      **03.2 HUMAN IN THE LOOP**
      **03.3 CONTRIBUTOR RECEPTIVITY TO IMPLEMENTING**
      **03.4 TRANSPARENCY IN USE OF AI**
      **03.5 EDUCATION AROUND AI**
      **03.6 EXISTING REGULATORY CULTURE**
      **03.7 LEADERSHIP BUY-IN**
**04 CHALLENGES TO SUCCESS**
      **04.1 ONGOING MONITORING**
      **04.2 ENFORCEMENT OF RESPONSIBILITY**
      **04.3 FUTURE-PROOFING**
      **04.4 SCALE AND SPEED OF WORK/COMMITMENT**
      **04.5 FINANCIAL TENSIONS**
      **04.6 EDUCATING STAFF ON POLICIES**
      **04.7 INTRODUCING MORE BUREACRACY**
      **04.8 EXPECTATION SETTING**
      **04.9 RESPONSIBLE AI AS MARKETING OR PERFORMATIVE**
      **04.10 SIZE OF COMPANY**
      **04.11 VENDOR AND PROCUREMENT CHALLENGES**
**05 FUTURE OPPORTUNITIES**
**06 RESOURCES**